\documentclass[letter]{aa}  
\usepackage{graphicx}
\usepackage{txfonts}

\usepackage{lipsum}
\usepackage{subcaption}         
\usepackage{lscape}             
\usepackage{placeins}           

\begin{document}

\title{Galactic warps: From cosmic noon to the current epoch}

\author{Vladimir P. Reshetnikov\inst{1,2}, Ilia V. Chugunov\inst{1}, Alexander A. Marchuk\inst{1,2}, Aleksandr V. Mosenkov\inst{3}, Matvey D. Kozlov\inst{2}, Sergey S. Savchenko\inst{1,2}, Dmitry I. Makarov\inst{1,4}, Aleksandra V. Antipova\inst{1,4}, Anastasia M. Sypkova\inst{1}
          }

\institute{Pulkovo Astronomical Observatory, Russian Academy of Sciences, St.Petersburg 196140, Russia \and 
Saint Petersburg State University, 7/9 Universitetskaya nab., St. Petersburg, 199034 Russia
\and
Department of Physics and Astronomy, N283 ESC, Brigham Young University, Provo, UT 84602, USA
\and
Special Astrophysical Observatory, Russian Academy of Sciences, Nizhnii Arkhyz 369167, Russia
}

\date{Received March , 2025; accepted , 2025}

\abstract
{Approximately half of all disk galaxies exhibit appreciable warps in both their stellar and H{\sc i} disks. The typical warp amplitude is small (a few degrees) and only becomes noticeable at the periphery of the galaxy disk. As a result, warps remain a relatively poorly studied phenomenon.} 
   {In this study, we investigate a large sample of distant edge-on galaxies (approximately 1,000 objects) in order to examine the frequency and characteristics of stellar disk warps up to a redshift of $z \sim 2$.
 }
   {For the selected galaxies, we used Hubble Space Telescope data from the Cosmic Evolution Survey field and JWST observations from the Cosmic Dawn Center Archive. We measured the properties of disk warps and investigated their evolution as a function of redshift.
}
   {Our results indicate a potential evolution in the observed frequency of strong S-shaped warps (with an amplitude greater than 4$^\circ$) in stellar disks as a function of redshift. At $z \approx 2$, the frequency of strong warps reaches approximately 50\%, while at $z \approx 0$, this fraction decreases to around 10--15\%. We attribute the observed evolution in the occurrence of strong warps to the changing frequency of galaxy interactions and mergers. If galaxy interactions represent one of the primary mechanisms responsible for the formation of warps, then the prevalence of vertical disk deformations should increase in tandem with the rising interaction and merger rate.
 }
   {}

\keywords{galaxies: structure -- galaxies: spiral -- galaxies: evolution
               }
\titlerunning{Galactic warps: From cosmic noon to the current epoch}
\authorrunning{V. P. Reshetnikov, I. V. Chugunov, A. A. Marchuk et al.}
\maketitle

\nolinenumbers

%

\section{Introduction}

The stellar and gaseous disks of spiral galaxies are known to be warped 
in the vertical direction, typically with an amplitude of a few degrees. This phenomenon has been observed in our Galaxy (e.g., \citealt{burke}; \citealt{kerr}; \citealt{djorg}) and in a significant number of other galaxies in the local Universe (\citealt{sanchez}; \citealt{r1995}; \citealt{degrijs}; \citealt{rc1998, rc1999}; \citealt{schwarz}; \citealt{garcia}; \citealt{sanchez2003}; \citealt{ann}; \citealt{guij}; \citealt{zee}). However, the mechanisms responsible for the origin and evolution of warps remain a subject of ongoing debate.

Several scenarios have been proposed to explain the formation and frequency of galactic warps. Among them are discrete modes of bending in a self-gravitating disk (\citealt{sparke}), interaction between the gaseous disk and extragalactic magnetic fields (\citealt{battaner}), misaligned dark halos (\citealt{dub}), galaxy interactions and satellite accretion (e.g., \citealt{huang}; \citealt{kim}; \citealt{sem}), and cosmic infall onto a disk galaxy (\citealt{shen}).
Despite these various hypotheses, none has achieved widespread acceptance. However, observations have indicated a correlation between the spatial environment and the generation of warps, as the most pronounced warps are typically found in denser environments (e.g., \citealt{rc1998, rc1999}; \citealt{schwarz}; \citealt{ann}; \citealt{ann2016}).

Most of our current understanding of bending deformations in galaxy disks is derived from studies of objects in the nearby Universe. The presence and characteristics of warps at earlier cosmic epochs remain largely unexplored. \citet{r2002} investigated a sample of edge-on galaxies in the Hubble Deep Fields North and South, reaching redshifts up to $z \sim 1$, and concluded that stellar disk warps were common in the past and appeared to have larger amplitudes. However, the results of \citet{r2002} were based on a relatively small sample (45 objects) and thus require confirmation using a larger dataset. In this paper, we combine the available data from the Hubble Space Telescope (HST) and JWST in order to investigate a significantly larger sample of approximately 1,000 edge-on galaxies, extending to redshifts up to $z \sim 2$, and study the optical warps of their disks.
Throughout this article, we adopt a standard flat $\Lambda$CDM cosmology with $\Omega_m$=0.3, $\Omega_{\Lambda}$=0.7, $H_0$=70 km\,s$^{-1}$\,Mpc$^{-1}$.


\section{The sample and data analysis}

The galaxies used in this study were selected from the HST/Advanced Camera for Surveys (ACS) COSMOS field, observed in the F814W filter, as well as from JWST data. The first sample comprises 950 edge-on galaxies. Detailed descriptions of the sample selection criteria and the databases used are provided in \citet{usachev}.

The second sample was constructed using data from the DAWN JWST Archive \citep[DJA;][]{valentino, brammer}, which contains images from seven fields observed by JWST: CEERS \citep{bagley}, JADES GOODS-S and GOODS-N \citep{rieke}, UDS-S and UDS-N, COSMOS-E, and COSMOS-W fields from PRIMER \citep{dunlop}. For these fields, we used publicly available DJA catalogs of objects selected using SExtractor++ \citep{bertin}. The selection criteria for edge-on galaxies included a photometric axis ratio of $b/a < 0.4$ in the F115W or F444W filters. We excluded galaxies with $B/T>0.5$ and with a major axis smaller than ten pixels (corresponding to 0.03 arcseconds per pixel). All remaining candidates were visually inspected by at least three authors, resulting in a final sample of 493 candidate galaxies. Additionally, we included edge-on galaxies identified by \citet{tsukui}, provided they were not already part of the selected sample.

For all the galaxies in the second sample, we retrieved redshift and stellar mass data from the DJA, where these values were derived through spectral energy distribution fitting using EAZY \citep{eazy}. The galaxies in the second sample are significantly smaller than those in the first sample, with an average redshift of $\langle z \rangle = 0.91$, and are distributed up to $z \approx 2-3$ (Fig.~1).

It is important to note that the second sample is inherently incomplete and does not encompass all edge-on galaxies detectable in the JWST observations of these fields. However, it is larger than the samples presented in \citet{lian} and \citet{tsukui}, primarily because we included galaxies with smaller major axis lengths compared to those considered in the other studies.

\begin{figure}
\centering
\includegraphics[width=8.5cm, angle=0, clip=]{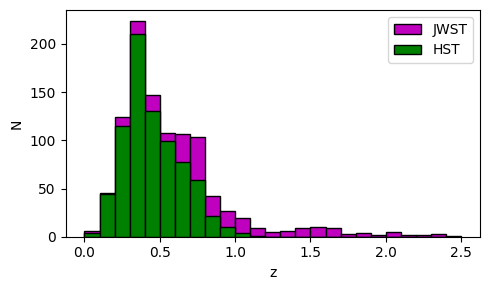}
\includegraphics[width=8.5cm, angle=0, clip=]{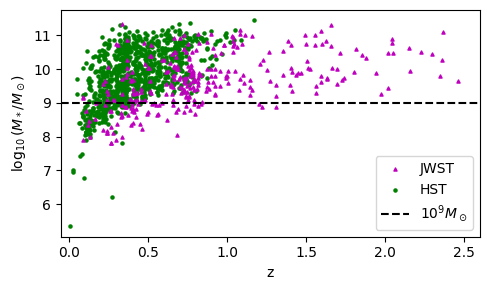}
\caption{Distribution of redshifts (top) and redshift dependence of stellar mass in solar units (bottom)
for the sample galaxies.}
\end{figure}

We began by measuring the center lines of the projected galaxy disks on the image plane, which in the absence of warps would appear straight. Following the approach described by \citet{r2016}, our method involves skeletonization of the isophotes. Essentially for each galaxy, we used multiple isophotes and selected the part of the image enclosed by each one. For each of these areas, we built its skeleton, which is a one-pixel wide line representing its general shape. We then averaged the locations of the skeletons and estimated the location of the center line. An example is shown in Fig.~2.

As a result, we obtained numerous measurements of the disk offset, $\Delta H$, from the disk plane at the center for different galactocentric radii, $r$. We fit the function $\Delta H(r)$ using two analytical approaches: a third-order polynomial and a piecewise linear function. While the latter method is less stable, it allowed us to define the onset of the warp as the break location, whereas we used the former method to estimate the warp angles.
For a given radius, $r$, we defined the warp angle, $\psi$, in a manner similar to \citet{r2016}; namely, it is the angle measured from the galactic center between the disk plane and the actual disk position at $r$ averaged over both sides of the center. However, the image depth often proved insufficient to accurately determine the disk shape near the isophotal radius, $R_{25}$, and thus prevented us from fully adopting this definition. Additionally, band-shifting effects and cosmological surface brightness dimming further complicated the measurements. To mitigate these challenges, we defined two specific warp angles: (i) warp angle $\psi_{4h}$ as measured at a distance of four exponential scale lengths of the disk ($4h$) from the center and (ii) warp angle $\psi_{e}$ measured at the location of the most distant $\Delta H(r)$ data point from the center that the method could capture, thus representing the outermost extent of the available disk shape measurements. Since not all galaxies in our sample have reliable $\Delta H(r)$ measurements extending to $4h$, we excluded $\psi_{4h}$ estimates for galaxies where $\Delta H(r)$ does not reach at least $3h$, as these values would rely on heavy extrapolation and are therefore unreliable. In the subsequent analysis, we primarily use $\psi_{e}$ as the main indicator of the warp angle, acknowledging that it inherently depends on the data's spatial extent.

\begin{figure}
	\centering
	\includegraphics[width=8.5cm, angle=0, clip=]{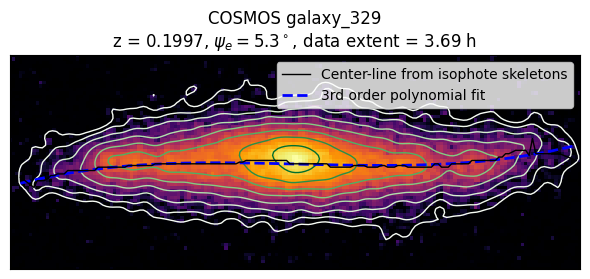}
	\caption{Example galaxy at $z = 0.20$ with a conspicuous S-shaped warp. Its isophotes, traced down to noise level, are displayed as green contours. The central line, determined from the isophote skeletons, is shown in black. For this object, the measured warp angle is $\psi_e = 5.3^\circ$.}
\end{figure}

We also measured the warp asymmetry as $\delta_\psi = \frac{|\psi_\text{l} - \psi_\text{r}|}{\psi_\text{l} + \psi_\text{r}}$, 
where $\psi_\text{l}$ and $\psi_\text{r}$ represent the absolute values of the warp angles on the left and right sides of the disk, respectively.
We distinguished between S-shaped and U-shaped warps by following the classification of \citet{rc1998}. A warp is considered S-shaped if the disk bends in opposite directions on either side of the galaxy center, while it is classified as U-shaped if the warp bends in the same direction on both sides.
Since the appearance of a U-shaped warp may result from a disk that is not truly edge-on (e.g., \citealt{rc1998}), we focused our statistical analysis primarily on S-shaped warps.

For galaxies where warps were measured using both F115W and F444W data, the mean difference between the warp angles $\psi_e^{F115W} - \psi_e^{F444W}$ is $0.2^\circ \pm 4.1^\circ$ for cases where $\psi_e > 4^\circ$ in both filters. This result indicates that we do not observe any systematic dependence of $\psi_e$ on wavelength (see, e.g., \citealt{zee}).

Figure\,1 (bottom) shows the distribution of the sample galaxies on the stellar mass–redshift plane. This figure reveals signs of observational selection. At higher redshifts, we primarily detect massive objects, while at lower redshifts, both low-mass and massive galaxies are observed.
To mitigate this selection bias, we restricted our analysis to galaxies with stellar masses of $M_* > 10^9\,M_{\odot}$ (the cutoff is indicated by the dashed line in Fig.\,1). As a result, our final sample comprises 1,027 edge-on galaxies, including 780 from the HST data and 247 from the JWST data, spanning redshifts up to $z \sim 2.5$.

\section{Results}

\subsection{Statistics of warps}

\begin{figure}
\centering
\includegraphics[width=8.5cm, angle=0, clip=]{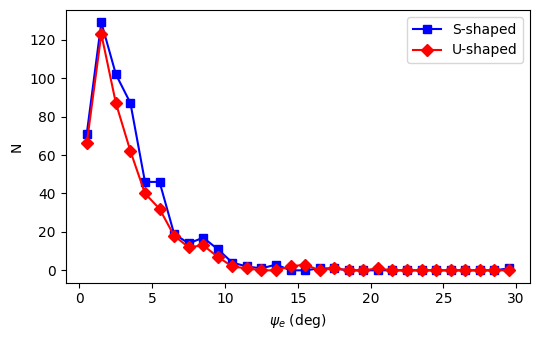}
\caption{Distribution of the warp angle $\psi_e$ (in degrees) for the S-shaped and U-shaped galaxies.}
\end{figure}

Figure 3 shows the distributions of the measured warp angles for the entire sample. The distributions for the two types of warps exhibit similar patterns, with global peaks at approximately $\psi_e \approx 2^\circ$. For the full sample, the total frequency of detected warps (demonstrating $\psi_e \geq 2^\circ$) is 62\%, comprising 35\% of S-shaped warps and 27\% of U-shaped warps. These values are comparable to those reported for the nearby Universe (e.g., \citealt{sanchez}; \citealt{rc1998}; \citealt{ann}; \citealt{zee}).

Measuring weak vertical disk deformations is challenging and subject to significant observational selection effects. Therefore, in the following discussion, we put an emphasis on the statistics of "strong" warps, which are defined as those with $\psi_e > 4^\circ$.

Figure 4 illustrates how the observed fraction of strong warps of different types varies with redshift. Despite the considerable scatter in the data, it is evident that the fraction of S-shaped warps increases monotonically up to $z \approx 2$, while the fraction of U-shaped warps remains approximately constant at around 10--20\%. At low redshifts, the observed fraction of strong S-shaped warps is approximately 10--15\% (Fig.\,4), which aligns with previous estimates (e.g., 11\% reported by \citealt{rc1998} and 15\% by \citealt{ann}). At $z \approx 2$, this fraction increases to about 50\% (Fig.\,4).
When we relax the constraint to $\psi_e > 2^\circ$, the overall trend remains unchanged: the fraction of S-shaped deformations continues to increase with redshift, while the fraction of U-shaped warps, after reaching a local maximum at $z \approx 1$, decreases toward $z \approx 2$. 

We observed no significant correlation between the $b/a$ of disks measured by decomposition and $z$, neither for the full sample nor individually for the S-shaped, U-shaped, or unwarped galaxies. Given that approximately 50\% of the spiral galaxies at $z \approx 2$ exhibit S-shaped warps with amplitudes greater than 4$^\circ$, it is likely that nearly all the spiral galaxies at $z \approx 2$ may be affected by substantial warping, considering the relationship between the observed and true warp fractions discussed by \citet{sanchez}.

\begin{figure}
\centering
\includegraphics[width=8.5cm, angle=0, clip=]{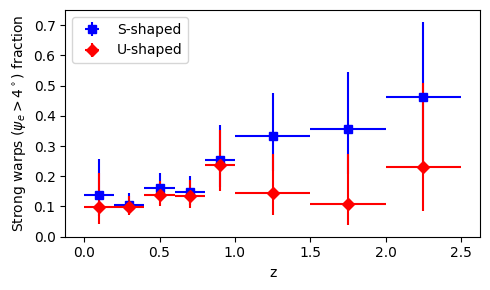}
\caption{Dependence of the fraction of S-shaped and U-shaped warps on redshift for $\psi_e > 4^\circ$ in multiple redshift bins. The horizontal error bars represent the redshift range, while the vertical ones limit the 95\% confidence interval.}
\end{figure}

\subsection{Characteristics of warps}

\begin{figure}
\centering
\includegraphics[width=8.5cm, angle=0, clip=]{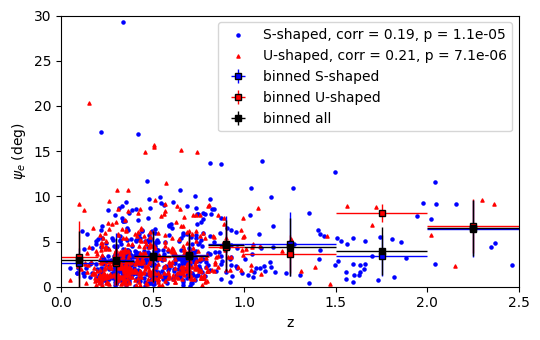}\\
\caption{Warp angle $\psi_e$ dependence on $z$. Squares with error bars show $\psi_e$ averaged in redshift bins. The horizontal error bars represent the redshift range of bins, while the vertical ones show the standard deviation.}
\end{figure}

Figure 5 shows the relationship between our measured bending angles and redshift for the entire sample of edge-on galaxies. Although the data points exhibit considerable scatter, a general trend of increasing warp angles with increasing redshift is evident. The Pearson correlation coefficient is modest (0.19 for the entire sample, 0.19 for S-shaped warps, and 0.21 for U-shaped), but given the number of points, it is statistically significant (the p-value is around $1 \cdot 10^{-5}$ for S-shaped and $7 \cdot 10^{-6}$ for U-shaped alone). It is important to note that the disks of galaxies at higher redshifts can typically be traced only to a greater surface brightness, which may limit the detection of some potential warps. Despite this limitation, the measured warp angle at the outermost point of the available data, $\psi_e$, is still higher for these high-redshift galaxies. It is likely that the actual warp angle would be even larger if the galaxy's periphery beyond the instrument detection limit were observable. This finding is consistent with the trend shown in Fig.\,4, where the fraction of strong warps increases with redshift.

For galaxies in different redshift ranges, we observed that the overall distribution of $\psi_e$ differs significantly.
When considering only S-shaped disks that are at least slightly warped ($\psi_e > 2^\circ$), the warp angle distributions can be well approximated by an exponential law. The scale parameter of this distribution is $2.1^\circ$ for nearby galaxies ($z < 0.35$) and $3.5^\circ$ for more distant objects ($z \geq 0.8$).

We performed artificial redshifting for a few bright galaxies from our sample (e.g., see Fig.\,5 in~\citealt{r2002}) to check for any possible changes in $\psi_e$ due to observational effects. We did this in the same way as described in Sec.~5.2.1 in \citet{pasa_spirals}. We conclude that measurements sometimes become unreliable if the extent of a galaxy's skeleton is too small. This condition can possibly lead to either an inability to capture the warp if it starts at the periphery of a galaxy or the appearance of a "false" warp due to isophotes becoming too influenced by noise. Thus, we additionally checked that the trend shown in Fig.\,5 remains significant if we remove galaxies with a small extent ($\le3h$) from the sample, and the trend is observed until $z \approx 2$.

Figure 6 shows the dependence of the projected starting point of stellar warps (expressed in units of the exponential scale length of the disk, $h$) on redshift. As can be seen in the figure, there is a clear relationship between the onset of bending and redshift: the beginning of the bend tends to decrease with increasing $z$ --- almost all points after $z>1$ lie below the average value. This is consistent with \citet{r2016}, whose authors noted that for nearby galaxies, stronger bends begin at a smaller (in fractions of $h$) distance from the center. Since we found that stronger warps are observed at large redshifts, the behavior portrayed in Fig.\,6 is expected. 

On average for the whole sample, warping begins at a distance of (1.6$\pm$1.0)\,$h$ from the galaxy center. Considering the variations in methodology for determining the onset of warping, this result is in reasonable agreement with previous findings for nearby galaxies. For instance, \citet{degrijs} reported a typical onset distance of (2.1$\pm$1.0)\,$h$ in the $I$ filter, while \citet{r2016} found a value of (2.15$\pm$0.5)\,$h$ for strong warps in the $i$ filter.

\begin{figure}
\centering
\includegraphics[width=8.5cm, angle=0, clip=]{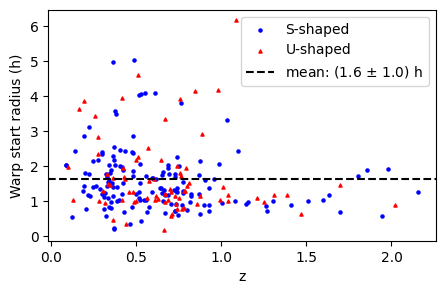}
\caption{Starting point of warp versus redshift (for $\psi_e > 4^\circ$).}
\end{figure}

The average value of warp asymmetry for our sample is $0.42 \pm 0.28$, which is nearly identical for both S-shaped ($0.41 \pm 0.28$) and U-shaped warps ($0.43 \pm 0.30$). These values are comparable to, or slightly higher than, the typical asymmetry observed in nearby galaxies (e.g., \citealt{sanchez2003}; \citealt{r2016}). We did not find any significant dependence of the warp asymmetry, $\delta_\psi$, on redshift. 

\section{Conclusions}

In this study, we have presented the results of analyzing the largest available sample (comprising approximately 1,000 objects) of distant edge-on galaxies. The data were collected from the HST and JWST archives. By analyzing galaxy images, we determined the characteristics of the vertical deformations of their stellar disks (warps) and investigated the dependence of these characteristics on redshift.

Similar to the local Universe, we found that the stellar disks of distant galaxies are frequently warped, indicating that warping is a characteristic feature of most distant disk galaxies. As redshift $z$ increases, the fraction of galaxies exhibiting strong warps also rises. At $z \approx 2$, approximately 50\% of galaxies display S-shaped warps with amplitudes greater than $4^\circ$. This is in stark contrast to the present-day Universe, where the fraction of such galaxies is only about 10--15\%.

A possible explanation for this significant evolution is the changing frequency of tidal interactions and mergers between galaxies as a function of redshift. Various methods for estimating the rate of galaxy interactions and mergers at earlier epochs --- such as statistics of close pairs and analyses of minor and major mergers across different redshifts --- consistently indicate that these processes were significantly more common in the past (e.g., \citealt{duncan}; \citealt{cons2022}; \citealt{ren}; \citealt{duan}).

Consequently, if the formation of S-shaped bending in a substantial fraction of galaxies is linked to external influences --- such as the accretion of matter, the merger of satellite galaxies, or tidal perturbations --- we would expect an increase in the occurrence of warps with redshift. Therefore, it is likely that the current statistics of stellar disk warps in distant galaxies reflect the more dynamic and interaction-rich evolutionary history of galaxies at earlier cosmic times.

Another important factor that may facilitate the formation of stellar warps at high redshifts is the increased  gas content of distant galaxies (e.g., \citealt{carilli}; \citealt{scoville}). 
Gas in spiral galaxies tends to be more widely distributed than stars, and it is more responsive to tidal perturbations and external accretion. Therefore, if a bending deformation has formed in the gas disk of a distant galaxy, it will manifest in the distribution of stars as a result of subsequent star formation.

\begin{acknowledgements}
This work was supported by the Russian Science Foundation (project no. 24-72-10084). Some of the data products presented herein were retrieved from the Dawn JWST Archive (DJA). DJA is an initiative of the Cosmic Dawn Center (DAWN), which is funded by the Danish National Research Foundation under grant DNRF140. This research is partially based on observations made with the NASA/ESA Hubble Space Telescope obtained from the Space Telescope Science Institute, which is operated by the Association of Universities for Research in Astronomy, Inc., under NASA contract NAS 5–26555. This work is based in part on observations made with the NASA/ESA/CSA James Webb Space Telescope. The data were obtained from the Mikulski Archive for Space Telescopes at the Space Telescope Science Institute, which is operated by the Association of Universities for Research in Astronomy, Inc., under NASA contract NAS 5-03127 for JWST.
\end{acknowledgements}


\begin{thebibliography}{}

\bibitem[Ann \& Park(2006)]{ann} Ann, H.B., \& Park, J.-C. 2006, New Astron., 11, 293

\bibitem[Ann \& Bae(2016)]{ann2016} Ann, H.B., \& Bae, H.J. 2016, JKAS, 49, 239

\bibitem[Bagley et al.(2023)]{bagley} Bagley, M.~B., Finkelstein, S.~L., Koekemoer, A.~M., et al.\ 2023, \apjl, 946, L12

\bibitem[Battaner, Florido \& Sanchez-Saavedra(1990)]{battaner} Battaner, E., Florido, E., \& 
Sanchez-Saavedra, M.L. 1990, A\&A, 236, 1

\bibitem[Bertin et al.(2020)]{bertin} Bertin, E., Schefer, M., Apostolakos, N., et al.\ 2020, Astronomical Data Analysis Software and Systems XXIX, 527, 461

\bibitem[Brammer(2023)]{brammer} Brammer, G.\ 2023, Zenodo

\bibitem[Brammer et al.(2008)]{eazy} Brammer, G.~B., van Dokkum, P.~G., \& Coppi, P.\ 2008, \apj, 686, 1503

\bibitem[Burke(1957)]{burke} Burke, B.F. 1957, AJ, 62, 90

\bibitem[Carilli \& Walter(2013)]{carilli} Carilli, C.L. \& Walter, F. 2013, \araa, 51, 105

\bibitem[Conselice et al.(2022)]{cons2022} Conselice, Ch.J., Mundy, C.J., Ferreira, L., \& Duncan, K. 2022,
ApJ, 940, 168

\bibitem[Chugunov et al. (2025)]{pasa_spirals} Chugunov I.~V., Marchuk A.~A., Mosenkov A.~V., 2025, PASA, 42, e029

\bibitem[de Grijs(1997)]{degrijs} de Grijs, R. 1997, PhD Thesis, Groningen University

\bibitem[Djorgovski \& Sosin(1989)]{djorg} Djorgovski, S., \& Sosin, C. 1989, AJ, 341, L13

\bibitem[Duan et al.(2024)]{duan} Duan, Q., Conselice, Ch.J., Qiong, L., et al. 2024, MNRAS, submitted (arXiv:2407.09472)

\bibitem[Dubinski \& Kuijken(1995)]{dub} Dubinski, J., \& Kuijken, K. 1995, ApJ, 442, 492

\bibitem[Duncan et al.(2019)]{duncan} Duncan, K., Conselice, Ch.J., Mundy, C., et al. 2019, ApJ, 876, 110

\bibitem[Dunlop et al.(2021)]{dunlop} Dunlop, J.~S., Abraham, R.~G., Ashby, M.~L.~N., et al.\ 2021, JWST Proposal. Cycle 1, 1837

\bibitem[Huang \& Carlberg(1997)]{huang} Huang, S., \& Carlberg, R.G. 1997, ApJ, 480, 503

\bibitem[Garc\'{i}a-Ruiz, Sancisi \& Kuijken(2002)]{garcia} Garc\'{i}a-Ruiz, I., Sancisi, R., \& Kuijken, K.
2002, A\&A, 394, 769

\bibitem[Guijarro et al.(2010)]{guij} Guijarro, A., Peletier, R.F., Battaner, E., et al. 2010,
A\&A, 519, A53

\bibitem[Kerr \& Hindman(1957)]{kerr} Kerr, F.J., \& Hindman, J.V. 1957, PASP, 69, 558

\bibitem[Kim et al.(2014)]{kim} Kim, J.H., Peirani, S., Kim, S., et al. 2014, ApJ, 789, 90

\bibitem[Lian \& Luo(2024)]{lian} Lian, J. \& Luo, L.\ 2024, \apjl, 960, L10

\bibitem[Ren et al. (2023)]{ren} Ren, J., Li, N., Liu, F.S., et al. 2023, ApJ, 958, 96

\bibitem[Reshetnikov(1995)]{r1995} Reshetnikov, V.P. 1998, Astron. Astrophys. Trans., 8, 31

\bibitem[Reshetnikov \& Combes(1998)]{rc1998} Reshetnikov, V., \& Combes, F. 1998, A\&A, 337, 9

\bibitem[Reshetnikov \& Combes(1999)]{rc1999} Reshetnikov, V., \& Combes, F. 1999, A\&AS, 138, 101

\bibitem[Reshetnikov et al.(2002)]{r2002} Reshetnikov, V., Battaner, E., Combes, F., 
\& Jim\'{e}nez-Vicente, J. 2002, A\&A, 382, 513

\bibitem[Reshetnikov et al.(2016)]{r2016} Reshetnikov, V., Mosenkov A., Moiseev A., Kotov S., \& Savchenko S. 2016, MNRAS, 461, 4233

\bibitem[Rieke et al.(2023)]{rieke} Rieke, M.~J., Robertson, B., Tacchella, S., et al.\ 2023, \apjs, 269, 16

\bibitem[S\'{a}nchez-Saavedra et al.(1990)]{sanchez} S\'{a}nchez-Saavedra, M.L., Battaner, E., \& Florido, E. 
1990, MNRAS, 246, 458

\bibitem[S\'{a}nchez-Saavedra et al.(2003)]{sanchez2003} S\'{a}nchez-Saavedra, M.L., Battaner, E., Guijarro, A. 
2003, A\&A, 399, 457

\bibitem[Schwarzkopf \& Dettmar(2001)]{schwarz} Schwarzkopf, U., \& Dettmar, R.-J. 2001, A\&A, 373, 402

\bibitem[Scoville et al.(2022)]{scoville} Scoville, N., Faisst, A., Weaver, J., et al. 2023, ApJ, 943, id.82

\bibitem[Semczuk et al.(2020)]{sem} Semczuk, M., \L{}okas, E.L., D'Onghia, E., et al. 2020, MNRAS, 498, 3535

\bibitem[Shen \& Sellwood(2006)]{shen} Shen, J., \& Sellwood, J.A. 2006, MNRAS, 370, 2

\bibitem[Sparke \& Casertano(1988)]{sparke} Sparke, L.S., \& Casertano, S. 1988, MNRAS, 234, 873

\bibitem[Tsukui et al.(2024)]{tsukui} Tsukui, T., Wisnioski, E., Bland-Hawthorn, J., et al.\ 2024, MNRAS, submitted (arXiv:2409.15909)

\bibitem[Usachev, Reshetnikov \& Savchenko(2024)]{usachev} Usachev, P.A., Reshetnikov, V.P., \& Savchenko, S.S.
2024, MNRAS, 529, L78

\bibitem[Zee et al.(2022)]{zee} Zee, W.-B. G., Yoon, S.-J., Moon, J.-S., et al. 2022, ApJ, 935, id.48

\bibitem[Valentino et al.(2023)]{valentino} Valentino, F., Brammer, G., Gould, K.~M.~L., et al.\ 2023, \apj, 947, 20

\end{thebibliography}
\end{document}